\documentclass[12pt,draftclsnofoot, onecolumn]{IEEEtran} 

\usepackage{amssymb}
\usepackage{amsthm}
\usepackage{graphicx}
\usepackage{array,color}
\usepackage{amsmath}
\usepackage{stfloats}
\usepackage{graphicx}
\usepackage{subfigure}
\usepackage{tabularx}
\usepackage{epsfig,epsf,color,balance,cite}
\usepackage{algorithmic}
\usepackage{algorithm}
\usepackage{bm}
\usepackage{textcomp}
\usepackage{color}
\usepackage{multirow}
\usepackage{cite}
\usepackage{enumerate}
\usepackage{cases}
\usepackage{color}
\usepackage{url}
\usepackage{epstopdf}

\begin{document}
	\title{UAV-Assisted and Intelligent Reflecting Surfaces-Supported Terahertz Communications}
	\author{\IEEEauthorblockN{Yijin Pan, Kezhi Wang, Cunhua Pan, Huiling Zhu and Jiangzhou Wang,~\IEEEmembership{Fellow,~IEEE} \vspace{-1em}}
	\thanks{Y. Pan are with the National Mobile Communications Research Laboratory, Southeast University, Nanjing 211111, China. She is also with School of Engineering and Digital Arts, University of Kent, UK. Email: panyj@seu.edu.cn, y.pan@kent.ac.uk.}
	\thanks{K. Wang is with the Department of Computer and Information Sciences, Northumbria University, UK. Email: kezhi.wang@northumbria.ac.uk.}
	\thanks{C. Pan is with the School of Electronic Engineering and Computer Science, Queen Mary, University of London, UK. Email: c.pan@qmul.ac.uk.}
	\thanks{J. Wang and H. Zhu are with the School of Engineering and Digital Arts, University of Kent, UK. Email: J.Z.Wang@kent.ac.uk, H.Zhu@kent.ac.uk.}}
	\maketitle

\begin{abstract}
	In this paper, unmanned aerial vehicles (UAVs) and intelligent reflective surface (IRS) are utilized to support terahertz (THz) communications.
	To this end, the joint optimization of UAV's trajectory, the phase shift of IRS, the allocation of THz sub-bands, and the power control is investigated to maximize the minimum average achievable rate of all the users.
	An iteration algorithm based on successive Convex Approximation with the Rate constraint penalty (CAR) is developed to obtain UAV's trajectory, and the IRS phase shift is formulated as a closed-form expression with introduced pricing factors.
	Simulation results show that the proposed scheme significantly enhances the rate performance of the whole system.
\end{abstract}


\section{Introduction}

Terahertz (THz) frequency band can provide abundant bandwidth to support the ultra-high data transmission for applications such as virtual reality, high-definition video broadcasting, etc. 
However, due to the ultra-high frequency, THz communications can be easily blocked by obstacles on the transmission path.
Recently, unmanned aerial vehicles (UAVs) are leveraged in wireless networks due to its highly flexible deployment with fully-controllable mobility. 
The wireless transmission supported by the low-altitude UAV can have a higher chance of line-of-sight (LoS) links with the ground user equipments (UEs).  
This makes the utilization of UAV to support THz communications a very promising solution.
Meanwhile, to serve multiple users, the intelligent reflective surface (IRS), also known as reconfigurable intelligent surface (RIS) has been proposed to help reconfigure wireless propagation channels.
By adjusting the phase shifts of the IRS's reflecting elements, the propagation environment can be significantly improved to help the communication of UEs with poor channel conditions \cite{9110849,9090356}. 
Therefore, a considerable performance gain can be obtained in the IRS-assisted and UAV-supported THz communication systems.

Current THz transmission approaches are mainly used for terrestrial communications. 
For example, the IRS was utilized in \cite{Chen.20198112019813,Pan.2020} to maximize the sum-rate performance of the THz communications.
The approach in \cite{Chaccour.2020} focused on utilizing the IRS to maintain reliable THz transmission. 
However, these contributions cannot be directly applied to the aerial scenario, where the UAV can be flexibly deployed for THz communications.
For the current contributions concerning IRS assisted-UAV transmission, single sub-band scenario was considered.
For instance, the UAV's trajectory and IRS's beamforming were jointly investigated in \cite{Li.2020}.
In fact, severer path loss peaks appear in the THz band and the locations of path loss peaks vary with the communication distance\cite{Han.2015, Han.2016}.
Thus, the THz band is composed of several distance-dependent sub-bands, and the current single sub-band scenario of IRS-assisted UAV solutions are not applicable.
Although there are a few contributions on orthogonal frequency division multiplexing (OFDM) communication in IRS-assisted UAV systems\cite{Lu.2020727,Yang.2019}, these approaches are based on the specially designed simplification of conventional channel model, which are not applicable to the THz communications.
As the path loss in the THz band is severely affected by the transmission distance, which highly relies on the trajectory of UAV.
Meanwhile, the sub-bands need to be intelligently selected for multiple UEs to avoid the path loss peaks.
Therefore, it is imperative to jointly optimize the trajectory of UAV, the phase shift of the IRS, along with the sub-band allocation to enhance the IRS-assisted and UAV-supported THz communication, which has not been studied in the current literature.

In this paper, we consider that the UAV supports THz communications and an IRS is deployed to help the transmission.
Our target is to maximize the minimum average rate among all UEs.
To address the formulated non-convexity problem, the optimization is decoupled into three subproblems, i.e., the trajectory of UAV, the phase shift of IRS, the THz sub-band allocation and the power control optimization.
In the UAV's trajectory optimization, we first utilize the successive convex approximation to simplify the complicated channel gain, and introduce penalties for the rate constraints to guarantee that the obtained objective function is non-decreasing.
For the phase shift optimization problem, the optimal solution can be obtained in a closed-form expression with introduced pricing factors. 
Finally, simulation results are provided to validate the convergence and effectiveness of the proposed algorithm.

\textit{Notation}: For a vector $\bm{x}$, $|\bm{x}|$ denotes its Euclidean norm.
$c$ represents the  light speed.
$\text{diag}(\bm{X})$ represents the vector that is obtained from the diagonal entries of matrix $\bm{X}$.

	\begin{figure}
		\centering
		\vspace{-1em}
		\includegraphics[width=0.6\textwidth]{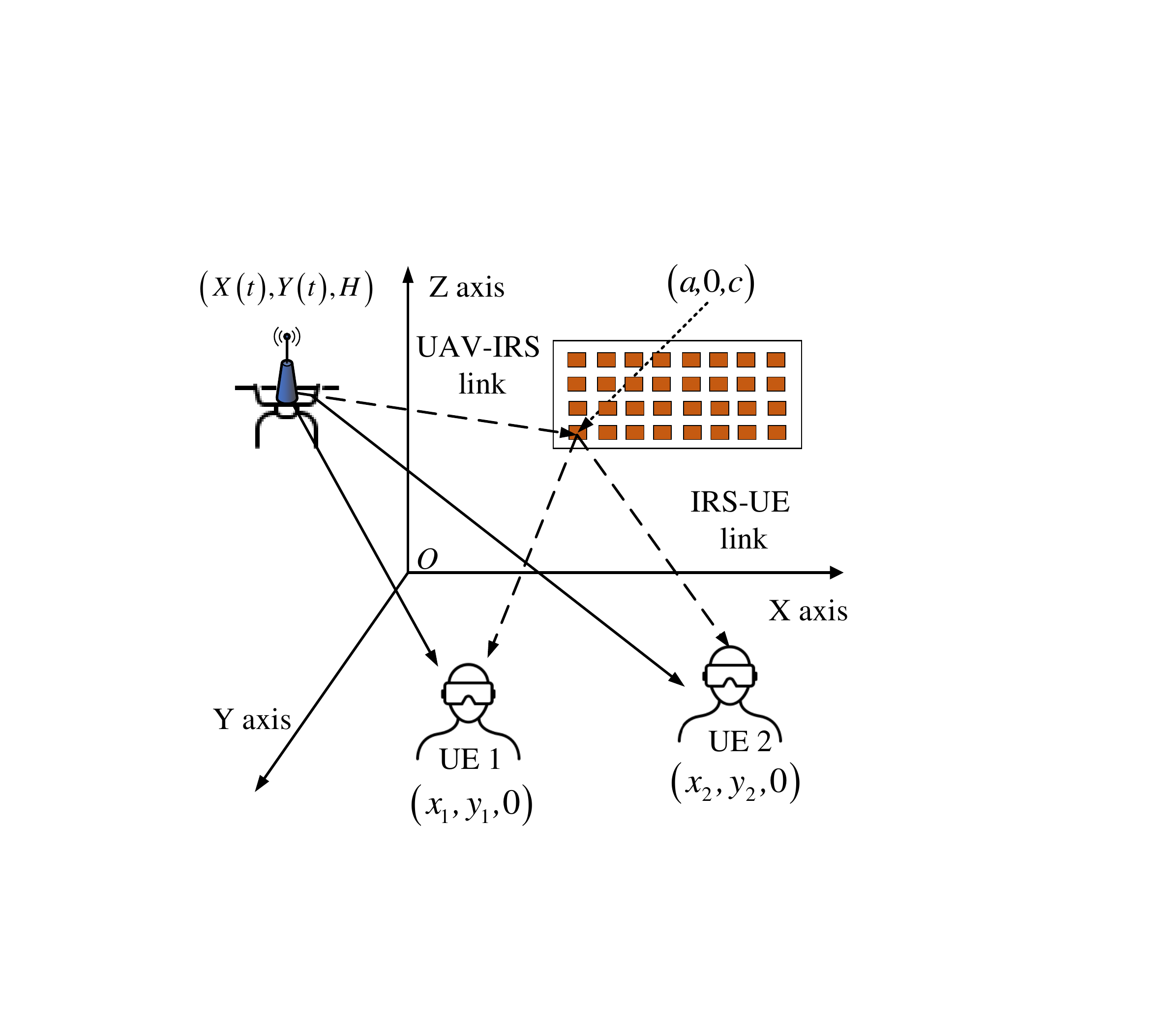}
		\vspace{-1em}
		\caption{The IRS-assisted and UAV-supported Thz communications.}
		\vspace{-1em}
		\label{fig1}
	\end{figure}

	\section{System Model and Problem Formulation}
	
	\newtheorem{proposition}{\textbf{Proposition}}
	
	Consider the downlink transmission of a UAV operating in THz frequency to serve $U$ UEs on the ground.
	The location of UE $u$ is denoted by $\bm{l}_u = [x_u,y_u,0]^T$.
	The UAV is assumed to fly at a fixed altitude $H$ above the ground within a total time of $T_s$. The total transmission time is divided into $T$ slots, and the location of the UAV and all channels are assumed to be unchanged within each time slot.
	The location of the UAV at time slot $t$ is denoted as $\bm{l}(t) = [X(t), Y(t), H]^T$.

	Consider there is an IRS deployed on a wall parallel to the XOZ plane, and the IRS is assumed to be a uniform planar array (UPA).
	Let $N_x$ and $N_z$ respectively represent the number of reflecting elements along the X-axis and Z-axis of the IRS, and the total number of reflecting elements is $N = N_xN_z$. 
	The separations between the elements along the X-axis and Z-axis are represented by $\delta_x$ and $\delta_z$, respectively.
	For the IRS,  the location of its first reflecting element is denoted as $\bm{l}_0 = [a,0,c]^T$.
	Then, the location of the $(n_x, n_z)$-th element of IRS is denoted by $[a +(n_x-1)\delta_x ,0,c+ (n_z-1)\delta_z]^T$. 
	Suppose that the reflection coefficients of all reflecting elements have the same amplitude value of $1$ but different phase shifts. 
	We use $\phi_{n}(t)$ to represent the phase shift of the $(n_x,n_z)$-th reflecting element at time slot $t$, where $n = n_z+ (n_x-1)N_z$.

	\subsection{Direct Transmission Links}
	
	The transmission distance between the UAV to UE $u$ at time slot $t$ is $d_u(t) = |\bm{l}(t) - \bm{l}_u|$.
	The UAV transmits signal through the THz band, which is affected by both free space spreading loss and the molecular absorption.
	As a result, the total bandwidth of the THz band is divided into several sub-bands to confront the frequency selective fading in THz band.
	Let $f_i$ denote the central frequency of the sub-band $i$, and the total number of sub-bands is denoted by $I$.
	The LoS channel gain from the UAV to UE $u$ on sub-band $i$ at time $t$ is denoted by 
	\begin{equation}
	h_{i,u}(t) = \left( \frac{c}{4 \pi f_i d_u(t)} \right) \exp\left({-j 2\pi f_i \frac{d_u(t)}{c}}\right) \exp\left({-\frac{1}{2}K(f_i)d_u(t)}\right),
	\end{equation}
	where $K(f_i)$ is the overall absorption coefficient of the transmission medium on sub-band $i$.
	
	\subsection{IRS-Assisted Transmission Links}

	The transmission vector from the UAV to the first element of the IRS can be represented as $\bm{r}_0(t) =\bm{l}_0- \bm{l}(t) = [a-X(t),-Y(t),c-H]^T$.
	The difference vector from the $(n_x,n_z)$-th element of the IRS to the first element of the IRS is $\Delta\bm{r}_{n_x,n_z}(t) =[(n_x-1)\delta_x,0,(n_z-1)\delta_z]^T$.
	Then, at time slot $t$, the relative phase difference on sub-band $i$ between the signal received at the first element and at the $(n_x,n_z)$-th element is 
	$\theta^i_{n_x,n_z}(t) = \frac{2\pi f_i}{c}\frac{\bm{r}_0(t)^T}{|\bm{r}_0(t)|}\Delta\bm{r}_{n_x,n_z}(t) 
	= \frac{2\pi f_i}{r(t)c}\left((a-X(t))(n_x-1)\delta_x\!+\!(c-H)(n_z-1)\delta_z\right)$,
	where $r(t) = |\bm{r}_0(t)|$.
	Then, the received array vector from UAV to the IRS at time $t$ on sub-band $i$ can be expressed as $\bm{e}_{i,r}(t) = [1, \cdots, \exp(-j \theta^i_{n}(t)), \cdots,\exp(-j \theta^i_{N}(t))]^T$, where the subscript ${n} = {n_z+ (n_x-1)N_z}$.
	
	The transmission vector from the first element of the IRS to UE $u$ is $\bm{r}_{u}= [(x_u-a),y_u, - c]^T$.
	Let $r_u = |\bm{r}_u|$. 
	Then, at time slot $t$, the relative phase difference on sub-band $i$ between the signal that is reflected by the first element and by the $(n_x,n_z)$-th element to UE $u$ is 
	$\vartheta^{i,u}_{n_x,n_z} = \frac{2\pi f_i}{c}\frac{\bm{r}_{u}^T}{|\bm{r}_{u}|}\Delta\bm{r}_{n_x,n_z}= \frac{2\pi f_i}{r_uc}((x_u-a)(n_x-1)\delta_x-c(n_z-1)\delta_z)$.	
	The transmit array vector from the IRS to UE $u$ at time $t$ on sub-band $i$ can be expressed as
	$\bm{e}_{i,u} = [1, \cdots, \exp(-j \vartheta^{i,u}_{n}), \cdots,\exp(-j \vartheta^{i,u}_{N})]^T$,
	where the subscript ${n} = {n_z+ (n_x-1)N_z}$.
	 		
	According to \cite{Tang.2019}, in the far field scenario, the cascaded channel gain of the UAV-IRS-UE $u$ link on sub-band $i$ is calculated as
	\begin{equation}
	 	\tilde{g}_{i,u}(t)= \left( \frac{c}{8 \sqrt{\pi^3}f_i  r_u r(t)} \right) \exp\left({-j 2\pi f_i \frac{(r(t)+r_u)}{c}}\right) \exp\left({-\frac{1}{2}K(f_i)(r(t)+r_u)}\right),
	\end{equation}
	where $K(f_i)$ is the absorption coefficient of the medium.	
	Then, the channel gain of the UAV-IRS-UE $u$ can be expressed as 
	$g_{i,u}(t)= \tilde{g}_{i,u}(t)\bm{e}_{i,r}(t) \bm{\Phi}(t){\bm{e}_{i,u}}^T$,
	 where $\bm \Phi(t)= \text{diag}(\exp(j\phi_{1}(t)), \cdots$ $,\exp(j\phi_{N}(t)))$, and $\exp(j\phi_{n}(t))$ represents the reflecting coefficient of the $n$-th reflecting element.
	
	\subsection{Problem Formulation}
	Let $B_i$ denote the bandwidth of sub-band $i$, and the power spectral density of the noise in sub-band $i$ is denoted by $S_N(f_i)$.
	The obtained transmission rate of UE $u$ on sub-band $i$ in time slot $t$ is given by 
	\begin{equation}
	R_{i,u}(t) =B_i \log \left( 1 + \frac{p_{i}(t)|h_{i,u}(t) + g_{i,u}(t)|^2 }{S_N(f_i) B_i}\right),
	\end{equation}
	where $p_{i}(t)$  is the transmit power on sub-band $i$ in the time slot $t$.
		
	In this problem, we first consider the following constraints for the UAV's trajectory $\{\bm{l}(t)\}$ as
	\begin{align}
	&C1:  \bm{l}(T) = \bm{l}(1) =\bm{\bar{l}}_1 , \\
	&C2:  |\bm{l}(t+1) - \bm{l}(t)| \leq V_{max} \frac{T_s}{T}, t = 1, \cdots, T-1.
	\end{align}
	The constraint $C1$ indicates that UAV needs to return to its initial location by the end of transmission period $T_s$, and the start point is fixed to the given location $\bm{\bar{l}}_1$. 
	The constraint $C2$ indicates that the maximum distance that the UAV can travel in each time slot is $V_{max} \frac{T_s}{T}$, where $V_{max}$ represent the maximum speed of the UAV.
	
	Define the binary variable $\alpha_{i,u}$  to indicate whether UE $u$ transmits on sub-band $i$ or not. 
	Assume that each sub-band is only allocated to one UE. Then, we have the following constraints as
	\begin{equation}
	C3: \alpha_{i,u}\in\{0,1\},\sum_{u =1}^U \alpha_{i,u}= 1, i\in\mathcal{I},u\in\mathcal{U}. 
	\end{equation}
	The transmit power $p_{i}$ is constrained by UAV maximum power $p_{max} $ as
	\begin{equation}
	C4: \sum_{i =1}^I p_{i}(t) \leq p_{max}.
	\end{equation}
	The phase shift of each reflecting element of the IRS varies within the range of $[0, 2\pi]$, we have
	\begin{equation} \label{C1st}
	C5:  0 \leq \phi_{n}(t) \leq 2\pi , 1\leq n \leq N. 
	\end{equation}

	Then, the problem can be formulated as
	\begin{equation}\label{Prom1}
	\begin{split}
		\underset{\underset{\{\bm{\Phi}(t)\},\{\alpha_{i,u}\}}{\{p_i\}, \{\bm{l}(t)\}}}{\text{max} }  
		&\ \underset{u}{\text{min}} \ R_u = \frac{1}{T}\sum_{t = 1}^T\sum_{i  \in \mathcal{I}} \alpha_{i,u}R_{i,u}(t)    \\
		\text{s.t.} & \ C1-C5,
	\end{split}
	\end{equation}
	where $R_u$ represents the average rate of UE $u$. Our target is to maximize the minimum average rate among all UEs by jointly optimizing the UAV's trajectory, THz sub-band allocation, the phase shift of IRS and the power allocation.
	By introducing auxiliary variables $R_{th}$ and $\{H^t_{i,u}\}$, Problem (\ref{Prom1}) can be equivalently reformulated as 
	\vspace{-0.5em}
	\begin{subequations}\label{Prom2}
		\begin{align}
		\underset{\underset{\{\bm{\Phi}(t)\},\{\alpha_{i,u}\}, \{p^i\}}{R_{th},\{H^t_{i,u}\},\{\bm{l}(t)\}}}{\text{max} }  
		&\quad \quad R_{th}  \label{obj2}  \\
		\text{s.t.} & \quad  \frac{1}{T}\sum_{t = 1}^T\sum_{i  \in \mathcal{I}} \alpha_{i,u}B_i \log \left( 1 + \frac{p_{i}(t)}{S_N(f_i) B_i}H^t_{i,u}\right) \geq  R_{th},u \in \mathcal{U},  \label{Prom2_st1}\\
		& \quad  |h_{i,u}(t)+ g_{i,u}(t)|^2 \geq \alpha_{i,u}H^t_{i,u},i\in\mathcal{I},u\in\mathcal{U},\label{Prom2_st2}\\
		& \quad C1-C5.\nonumber
		\end{align}
	\end{subequations}

\section{Solution Analysis}
As the formulated problem (\ref{Prom2}) is non-convexity, we decouple the optimization  into three subproblems: optimization of UAV's trajectory, optimization of the  IRS's phase shift, the THz sub-band allocation and the power control problem.

\subsection{Trajectory Optimization}

Given the THz sub-band allocation, the phase shift of IRS and the power control, the UAV's trajectory is optimized in this subsection.
The trajectory optimization is challenging due to the complicated expression of the overall channel gain, which is 
\begin{multline} \label{Hsum}
H_{i,u}(\bm{l}(t)) =|h_{i,u}(t)+ g_{i,u}(t)|^2 = \left( \frac{A_i}{d_u(t)} \right)^2\exp\left({-K_id_u(t)}\right) + \\
\left( \frac{B_{i,u}}{r(t)} \right)^2 \exp\left({-K_ir(t)}\right)C_{i,u}(\bm{l}(t))
+ \frac{2D_{i,u} (\bm{l}(t))}{d_u(t)r(t)}  \exp\left({-\frac{1}{2}K_i(r(t)+d_u(t))}\right),
\end{multline}
where $A_i$, $B_{i,u}$, $K_i$, $C_{i,u}(\bm{l}(t))$, and $D_{i,u} (\bm{l}(t))$ are respectively given by
\begin{align}
&A_i = \frac{c}{4\pi f_i}, B_{i,u} = \left( \frac{c}{8 \sqrt{\pi^3}f_i  r_u} \right) \exp\left({-\frac{1}{2}K(f_i)r_u}\right),  K_i = K(f_i),\\
&C_{i,u}(\bm{l}(t)) =\left|\bm{e}_{i,r}(t) \bm{\Phi}{\bm{e}_{i,u}}^T\right|^2,  \label{C}\\
&D_{i,u} (\bm{l}(t))= A_iB_{i,u}\Re\left\{ \exp\left({-j 2\pi f_i \frac{(r(t)+r_u) -d_u(t)}{c}}\right)\bm{e}_{i,r}(t) \bm{\Phi}{\bm{e}_{i,u}}^T  \right\}. \label{D}
\end{align}
It is observed that $C_{i,u}(\bm{l}(t))$ and $D_{i,u} (\bm{l}(t))$ consist of many periodic cosine patterns with respect to the sub-bands and UE indexes, which make the channel gain given in (\ref {Hsum}) difficult to handle.
However, according to $C2$, with a sufficiently short time slot, the change of UAV’s locations in a time slot is very small. 
As a result, we can first regard $C_{i,u}(\bm{l}(t))$ and $D_{i,u}(\bm{l}(t))$ as
constants and denote them as
$C_{i,u}(\bm{l}(t)) \triangleq C_{i,u}^t$ and $D_{i,u}(\bm{l}(t)) \triangleq D_{i,u}^t$, respectively.
In addition, we define functions
$f_i(x) = \frac{1}{x^2} \exp(-K_i x)$ and $q_i(x,y) = \frac{1}{xy} \exp\left(-\frac{K_i}{2} (x+y)\right)$.
Then, the channel gain in (\ref{Hsum}) can be approximated as
\begin{equation}\label{ApproxH}
\hat{H}_{i,u}(\bm{l}(t))
= A_i^2 f_i(d_u(t)) + C_{i,u}^t{B^2_{i,u}} f_i(r(t))+ 2D_{i,u}^tq_i (d_u(t),r(t)) .
\end{equation}
Then, by checking the second-order derivative of $f_i(x)$ with respect to $x$, we have
\begin{equation}
\frac{\text{d}^2 f_i(x)}{\text{d} x^2} = \frac{(K_ix +2)^2 +2}{x^4}\exp(-Ax) >0.
\end{equation}
In addition, the Hessian matrix of $q_i(x,y)$ is given by
\begin{equation}
\triangledown^2 q_i(x,y) 
= \frac{1}{xy}\exp\left(-\frac{K_i}{2}(x+y)\right)
\left[\begin{matrix}
\frac{(\frac{K_i}{2}x+1)^2+1}{x^2} & \frac{(\frac{K_i}{2}x+1)(\frac{K_i}{2}y+1)}{xy}\\
\frac{(\frac{K_i}{2}x+1)(\frac{K_i}{2}y+1)}{xy} & \frac{(\frac{K_i}{2}y+1)^2+1}{y^2} 
\end{matrix}\right].
\end{equation}
It can be readily verified that the determinant of $\triangledown^2 q_i(x,y)$ is positive, so that $\triangledown^2 q_i(x,y)$ is positive definite.
As a result, $f_i(x)$ and $q_i(x,y)$ are convex functions with respect to $x$ and $(x,y)$, respectively.

By substituting $|h_{i,u}(t)+ g_{i,u}(t)|^2$ with $\hat{H}_{i,u}(\bm{l}(t))$ in (\ref{ApproxH}), (\ref{Prom2_st2}) are verified to be non-convex constraints.
Then, the successive convex approximation (SCA) is adopted to 
reformulate constraint (\ref{Prom2_st2}) by utilizing the first-order Taylor approximations of
$f_i(x)$ and $q_i(x,y)$.
At a given location $\bm{l}'(t) = [X'(t), Y'(t), H]$, let us define 
\begin{equation}
C_{i,u}^{t'} \triangleq C_{i,u}(\bm{l}'(t)), D_{i,u}^{t'} \triangleq D_{i,u}(\bm{l}'(t)), 
d_u^{t'} \triangleq |\bm{l}'(t) - \bm{l}_u|,{r^t}'\triangleq|\bm{l}_0- \bm{l}'(t)|.
\end{equation}
Then, constraint (\ref{Prom2_st2}) can be approximated as
\begin{equation}\label{Prom3_Conv}
\zeta^t_{i,u}{d_u^t}+  \psi^t_{i,u} {r^t} + const_{i,u}^t \geq \alpha_{i,u}H^t_{i,u} ,
\end{equation}
where 
\begin{align}
&\zeta^t_{i,u} = \left(A_i^2\frac{\text{d} f_i }{\text{d} x}(d_u^{t'}) + 2D^{t'}_{i,u} \frac{\partial q_i }{\partial x}(d_u^{t'},r^{t'})\right).\\
&\psi^t_{i,u} = \left(C^{t'}_{i,u}{B^2_{i,u}} \frac{\text{d} f_i }{\text{d} x}(r^{t'}) +  2D^{t'}_{i,u} \frac{\partial q_i }{\partial y}(d_u^{t'},r^{t'})\right) \\
&const_{i,u}^t = A_i^2 f_i(d_u^{t'}) + C^{t'}_{i,u}{B^2_{i,u}} f_i(r^{t'})+2D^{t'}_{i,u} q_i (d_u^{t'},r^{t'}) - \psi^t_{i,u}r^{t'} - \zeta^t_{i,u}d_u^{t'}.
\end{align}
Note that the first-order derivatives of $f_i(x)$ and $q_i(x,y)$ are respectively given by 
\begin{align}
&\frac{\text{d} f_i }{\text{d} x}(x) = - \frac{K_i x +2}{x^3} \exp(-K_ix),  \\
&\frac{\partial q_i }{\partial x}(x,y) = \frac{-(\frac{K_i}{2} x+1)}{x^2y} \exp\left(-\frac{K_i}{2}(x+y)\right).
\end{align}

It is verified that $f_i(x)$ is a decreasing function with respect to $x$, and $q_i(x,y)$ is a decreasing function with respect to $x$ and $y$, respectively.
Then, given $\bm{L}_{-t} = [\bm{l}(1),\cdots,\bm{l}(t-1),\bm{l}(t+1),\cdots,\bm{l}(T)]$, and a given reference location $\bm{l}'(t) = [X'(t), Y'(t), H]$, the UAV's location $\bm{l}(t) = [X(t), Y(t), H]^T$ at time slot $t$ can be determined by solving the following optimization problem
\begin{subequations}\label{Prom3}
	\begin{align}
	\underset{\bm{l}(t), d^t_u,r^t,\{H^t_{i,u}\}}{\text{max} }  
	&\quad R_{th}  \label{obj3}  \\
	\text{s.t.}  & \quad \sum_{i  \in \mathcal{I}} \alpha_{i,u}B_i \log \left( 1 + \frac{p_{i}(t)}{S_N(f_i) B_i}H^t_{i,u}\right) + R_{u}^{-t} \geq  T R_{th},u\in\mathcal{U},  \label{Prom3_st1} \\
	&\quad |\bm{l}(t) - \bm{l}_u|\leq d_u^t,u\in\mathcal{U},\label{Prom3_st3}\\ 
	&\quad |\bm{l}_0- \bm{l}(t)| \leq r^t,\label{Prom3_st4} \\
	&\quad (\ref{Prom3_Conv}),C1,C2, \nonumber
\end{align}
\end{subequations}
where $\{d_u^t\}$ and $r^t$ are auxiliary variables, and $R_{u}^{-t} = \sum_{s \neq t }^T\sum_{i  \in \mathcal{I}}\alpha_{i,u}R_{i,u}(s)$.
Problem (\ref{Prom3}) is a convex problem, and it can be readily solved by the standard solvers, such as the CVX.

However, note that the channel gain given in (\ref{ApproxH}) is a simplified version of (\ref{Hsum}).
Let $(\bm{\hat{l}}(t),\hat{H}^t_{i,u})$ denote the optimal solution to Problem (\ref{Prom3}).
According to (\ref{ApproxH}), we have $\hat{H}^t_{i,u} \neq H_{i,u}(\bm{\hat{l}}(t))$, so that the constraint in (\ref{Prom3_st1}) may not be satisfied with $(\bm{\hat{l}}(t),H_{i,u}(\bm{\hat{l}}(t)))$.
Define
\begin{equation}
\Delta_u = \sum_{i  \in \mathcal{I}} \alpha_{i,u}B_i \log \left( 1 + \frac{p_{i}(t)}{S_N(f_i) B_i}H_{i,u}(\bm{\hat{l}}(t))\right) + R_{u}^{-t} - T R_{th}. 
\end{equation}
If $\Delta_u < 0$, the obtained $(\bm{\hat{l}}(t),H_{i,u}(\bm{\hat{l}}(t)))$ degrades the average achievable rate $R_u$ of UE $u$.
Let $\mathcal{U}_R$ denotes the set of UEs with $\Delta_u < 0$.
To guarantee  that $R_u \geq R_{th}$ for all UEs in $\mathcal{U}_R$, more strict constraints are introduced by defining
\begin{equation}
\hat{R}_{u}^{-t} = R_{u}^{-t} + \Delta_{u} , \forall u \in \mathcal{U}_R. 
\end{equation}

Correspondingly, the constraint in (\ref{Prom3_st1}) is updated by substituting ${R}_{u}^{-t}$ with $\hat{R}_{u}^{-t}$ for all $u \in \mathcal{U}_R$. 
Then, by continuously solving the updated Problem (\ref{Prom3}) with the modified rate constraints of the UEs in $\mathcal{U}_R$, the obtained solution $(\bm{\hat{l}}(t),H_{i,u}(\bm{\hat{l}}(t)))$ will not degrade the average achievable rate of UE $u$.
However, it is worth pointing out that the updated problem may not be feasible, as the change of UAV’s location in a time slot is very small due to constraint $C2$.
In this case, no new UAV's position that can improve the current objective value $R_{th}$ can be found, i.e., the UAV should stay in the current position $\bm{l}'(t)$.
Overall, the above analysis can be summarized as the following CAR Algorithm \ref{alg1} to optimize the UAV's trajectory.

\begin{algorithm}
	\caption{ Successive Convex Approximation with Rate constraint penalty (CAR)}
	\begin{algorithmic}[1]\label{alg1}
		\STATE Initialize the convergence precision $\bm{\varsigma}$, trajectory $\{\bm{l}(t)^{0}\}$ and the iterative number $m=0$.
		\REPEAT
		\FOR{$t = 1, \cdots, T$}
		\STATE Initialize $\bm{l}'(t) = \bm{l}(t)^{m}$ and $\mathcal{U}_R = \mathcal{U}$;
		\REPEAT  
		\STATE Obtain $\bm{\hat{l}}(t)$ by solving Problem (\ref{Prom3});  
		\STATE Update $\mathcal{U}_R$, replace $R_{u}^{-t}$ with $\hat{R}_{u}^{-t}$ for all $u \in \mathcal{U}_R$;
		\UNTIL $\mathcal{U}_R = \varnothing$ or Problem (\ref{Prom3}) is not feasible;
		\IF{$\mathcal{U}_R = \varnothing$}
		\STATE Update $\bm{l}(t)^{m} =\bm{\hat{l}}(t)$;
		\ENDIF
		\ENDFOR
		\UNTIL  $|\bm{l}(t)^{m+1} -\bm{l}(t)^{m} |\leq \bm{\varsigma}, \forall t $ .
	\end{algorithmic}
\end{algorithm}

\subsection{Phase Shift Optimization}
Given the THz sub-band allocation, the power control, and the UAV's trajectory, we first reformulate (\ref{Hsum}) as 
\begin{equation}\label{Hphi}
|h_{i,u}(t)+ g_{i,u}(t)|^2 = G_{i,u}(t) + F_{i,u}(t)|\bm{v}_{i,u}(t) \bm{\phi}(t)|^2 + \Re\{Q_{i,u}(t)\bm{v}_{i,u}(t) \bm{\phi}(t)\},
\end{equation}
where 
\begin{align}
&G_{i,u}(t)= \left( \frac{A_i}{d_u(t)} \right)^2\exp\left({-K_id_u(t)}\right) ,F_{i,u}(t)= \left( \frac{B_{i,u}}{r(t)} \right)^2 \exp\left({-K_ir(t)}\right),\\
&Q_{i,u}(t) = \frac{2A_iB_{i,u}}{d_u(t)r(t)}  \exp\left({-\frac{1}{2}K_i(r(t)+d_u(t))}-j 2\pi f_i \frac{(r(t)+r_u) -d_u(t)}{c}\right),\\
&\bm{v}_{i,u}(t) = [1, \cdots,\exp(-j( \vartheta^{i,u}_{N}+\theta^i_{N}(t)))]^T, \bm{\phi}(t) = \text{diag}(\bm{\Phi}(t)).
\end{align}

Define $s_{i,u}(t)=\bm{v}_{i,u}(t) \bm{\phi}(t)$, and it is observed that the right hand side of (\ref{Hphi}) can be formulated as a convex function with respect to $s_{i,u}(t)$ as 
\begin{equation}
U({s}_{i,u}(t)) = G_{i,u}(t) + F_{i,u}(t)s_{i,u}(t)s^*_{i,u}(t) + \frac{1}{2}Q_{i,u}(t)s_{i,u}(t) + \frac{1}{2}Q_{i,u}(t)^*s^*_{i,u}(t).
\end{equation}

At a given point $\tilde{s}_{i,u}(t)=\bm{v}_{i,u}(t)\bm{\tilde{\phi}}(t)$,  we have the following inequality as 
\begin{equation}\label{Usinq}
U({s}_{i,u}(t)) \geq  U(\tilde{s}_{i,u}(t))  +  \Re\left\{ \left(2F_{i,u}(t)\tilde{s}^*_{i,u}(t) + Q_{i,u}(t)  \right)(\bm{v}_{i,u}(t) \bm{\phi}(t) -\tilde{s}_{i,u}(t))\right\}. 
\end{equation}

The phase shift of IRS can be optimized by solving a sequence of problems as
\begin{subequations}\label{Prom5}
	\begin{align}
	\underset{0 \leq \phi_{n}(t) \leq 2\pi}{\text{max} }  
	&\quad \quad R_{th}  \label{obj5}  \\
	\text{s.t.} & \quad   \alpha_{i,u}(\Re\left\{ \Upsilon_{i,u}(t)\bm{v}_{i,u}(t) \bm{\phi}(t)\right\}  - \chi_{i,u}(t)) \geq  0,
	\end{align}
\end{subequations}
where  $\Upsilon_{i,u}(t) = \left(2F_{i,u}(t)\tilde{s}^*_{i,u}(t) + Q_{i,u}(t)\right)$ and $\chi_{i,u}(t) = \Re\left\{\Upsilon_{i,u}(t)\tilde{s}_{i,u}(t)\right\}  -U(\tilde{s}_{i,u}(t)) + H^t_{i,u} $.

Although Problem (\ref{Prom5}) is still nonconvex, as proved in \cite{Pan.2020}, its globally optimal solution can be found by the following procedure.
First, a penalty is added in the objective function with the pricing factors $\{\rho_{i}(t)\}$. Then, the phase shift can be determined by 
\begin{subequations}\label{Prom6}
	\begin{align}
	\underset{ 0 \leq \phi_{n}(t) \leq 2\pi}{\text{max} }  
	&\quad \quad R_{th} + \sum_{i  \in \mathcal{I}} \rho_{i}(t)\sum_{u =1}^U\alpha_{i,u}(\Re\left\{ \Upsilon_{i,u}(t)\bm{v}_{i,u}(t) \bm{\phi}(t)\right\}  - \chi_{i,u}(t) ).                                         \label{obj6}
	\end{align}
\end{subequations}
The solution to Problem (\ref{Prom6}) is given by 
\begin{equation}
 \phi_{n}(t)  = \angle(\Xi_n(t)),
 \end{equation}
 and $\Xi_n(t)$ is the $n$-th element of $\bm{\Xi}(t)$, which is
\begin{equation}
\bm{\Xi}(t) =  \sum_{i  \in \mathcal{I}} \rho_{i}(t)\alpha_{i,u} (\Re\left\{ \Upsilon^*_{i,u}(t)\bm{v}^*_{i,u}(t)\right\}.
\end{equation}
To reduce the penalty, the pricing factors are updated by sub-gradient descent method.
The $\{\rho_{u,i}(t)^{(s)}\}$ in the $s$-th iteration is updated as
\begin{align}
\!\!\!\!\!\rho_{u,i}^{(t)} \!=\! \left[\rho_{u,i}^{(t-1)}\! - \!\tau_{u,i}^{(t)}\left(\sum_{u =1}^U\alpha_{i,u}(\Re\left\{ \Upsilon_{i,u}(t)\bm{v}_{i,u}(t) \bm{\phi}(t)\right\}  - \chi_{i,u}(t) ) \right)\!\right]^+ \!\!\!,\label{rhokt} 
\end{align}
where $[a]^+ = \max\{0,a\}$, $\tau_{u,i}^{(t)}$ is the positive step-size in the $t$-th iteration.

\subsection{THz Sub-Band Allocation and Power Control}	

Given UAV's trajectory, and the phase shift of the IRS, the THz sub-band allocation and power control are optimized as
\begin{subequations}\label{Prom7}
	\begin{align}
	\underset{\{\alpha_{i,u}\}, \{p^i(t)\},R_{th}}{\text{max} }  
	&\quad \quad R_{th}  \label{obj7}  \\
	\text{s.t.} & \quad  \frac{1}{T}\sum_{t = 1}^T\sum_{i  \in \mathcal{I}} \alpha_{i,u}B_i \log \left( 1 + p_{i}(t) \frac{H^t_{i,u}}{S_N(f_i) B_i}\right) \geq  R_{th},  \label{Prom7_st1}\\
	& \quad C3,C4. \nonumber
	\end{align}
\end{subequations}
Introducing transformation $x_{u,i}(t) = \alpha_{u,i}p_{i}(t)$, the above Problem (\ref{Prom7}) can be solved by using the dual-based method given in \cite{Wong.1999}.

In summary, based on the above analysis, we propose the following Algorithm \ref{alg2} to solve the original Problem (\ref{Prom2}).
	\begin{algorithm}
	\caption{Algorithm to Solve Problem (\ref{Prom2})}
	\begin{algorithmic}[1]\label{alg2}
		\STATE Initialize UAV's trajectory $\bm{l}^0(t)$ and the phase shift of the IRS $\bm{\Phi}^0(t)$.
		\STATE Initialize the convergence precision $\sigma$ and the iterative number $m=0$.
		\REPEAT
		\STATE Calculate $\alpha_{u,i}^{(m+1)}$, ${p_i(t)}^{(m+1)}$ and $R_{th}^{(m+1)}$ by solving Problem (\ref{Prom7});
		\STATE Calculate $\{\bm{l}(t)^{(m+1)}\}$ according to CAR Algorithm;
		\STATE Initialize precision $\bm{\theta}$, $\bm{\phi}(t)^{(0)} = \text{diag}(\bm{\Phi}(t)^{m})$ and iterative number $n=0$;
		\REPEAT
		\STATE Formulate Problem (\ref{Prom6}) with $s_{i,u}(t)^{n}=\bm{v}_{i,u}(t) \bm{\phi}(t)^{(n)}$;
		\STATE Calculate $\bm{\phi}(t)^{(n+1)}$ by solving Problem (\ref{Prom6}); 
		\UNTIL  $|\bm{\phi}(t)^{(n+1)}-\bm{\phi}(t)^{(n)}|\leq \bm{\theta}$;
		\STATE  Obtain  $\bm{\Phi}(t)^{m+1}$ according to  $\bm{\phi}(t)^{(n+1)}$;
		\UNTIL $|R_{th}^{(m+1)}-R_{th}^{(m)}|\leq \sigma$.
	\end{algorithmic}
\end{algorithm}

\section{Simulation Results}	

Simulation results are presented in this section to evaluate the proposed scheme.
In our simulation, the transmission frequency is $200$-$400$ GHz, the bandwidth of each sub-band is $10$ GHz,
and the molecular absorption coefficient $K(f_i)$ is generated according to \cite{Boulogeorgos.20186252018628}.
The IRS is located in the middle of X-axis with the hight of $2$ m, $N_x = 8$, $N_z = 10$, and $\delta_x = \delta_z = 5$ mm.
The flying height of the UAV is $10$ m, maximum transmit $p_{max}$ is $1$ W, maximum speed $V_{max} = 2$ m/s, transmission duration $T_s =2$ min, and the number of time slots is $T=50$.
For comparison, we consider three different algorithms: {1) The sub-band allocation and power control are randomly selected during the iterations and labeled as ``PwCh fixed''; 2) The phase shift of the IRS is randomly generated during the iterations and labeled as ``Theta fixed''; 3) The trajectory of the UAV is fixed as the initialized one during the iterations and labeled as ``Traject fixed''.

\begin{figure}
		\begin{minipage}[t]{0.49\textwidth}
		\centering
		\vspace{-1em}
		\includegraphics[width=1\textwidth]{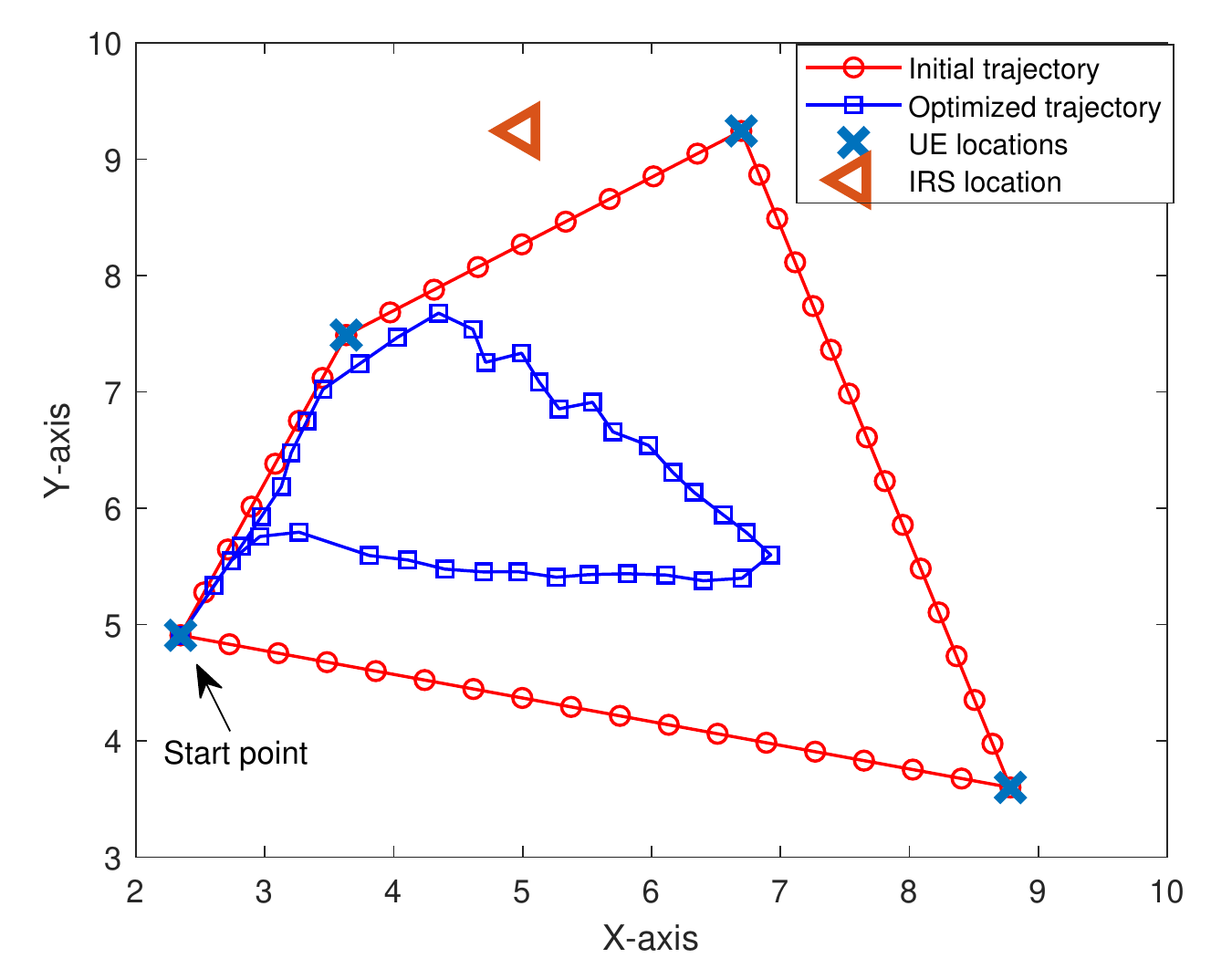}
		\vspace{-3em}
		\caption{Converged trajectory of the UAV.}
		\vspace{-2em}
		\label{fig2}
	\end{minipage}
	\begin{minipage}[t]{0.02\textwidth}
		\centering
		\vspace{-1em}
		\includegraphics[width=1\textwidth]{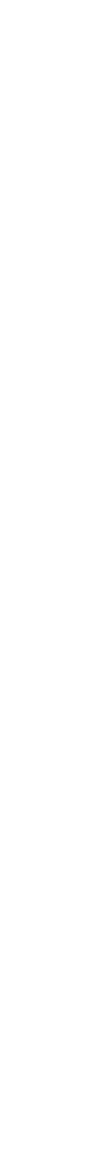}
		\vspace{-3em}
	\end{minipage}
	\begin{minipage}[t]{0.49\textwidth}
	\centering
	\vspace{-1em}
	\includegraphics[width=1\textwidth]{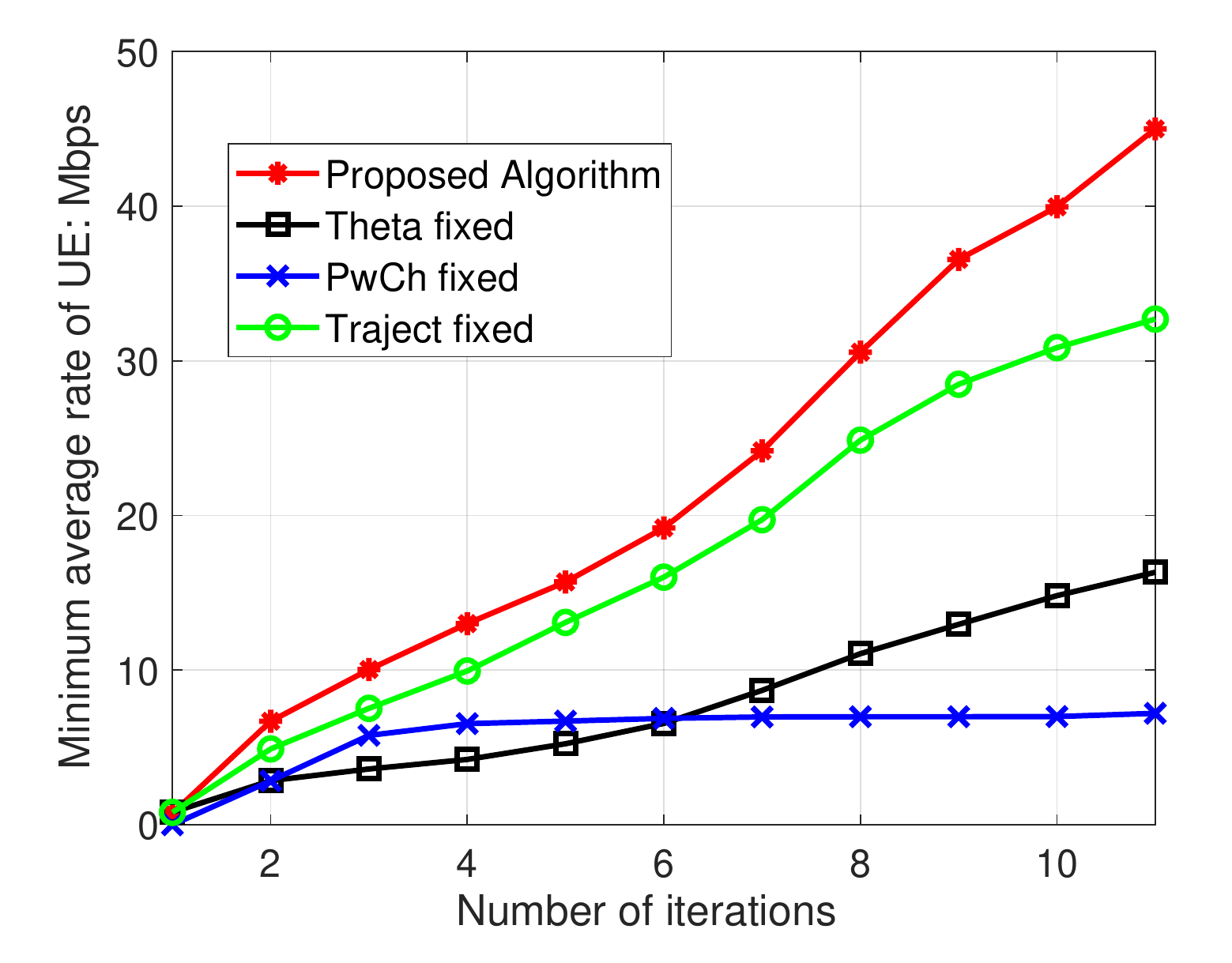}
	\vspace{-3em}
	\caption{Convergence performance of different algorithms.}
	\vspace{-2em}
	\label{fig3}
\end{minipage}
\end{figure} 

Fig. \ref{fig2} illustrates the optimized trajectory, which is a compromise between the distances to UEs and the distance to the IRS.
It is observed that the optimized trajectory requires the UAV to move a much smaller distance than that of the initial trajectory, which is attributed to the jointly optimized phase shift of IRS, the THz channel allocation, and the power control.
This also implies that UAV can save energy for the flight and the proposed algorithm fully exploits the benefit brought by the joint design of UAV's trajectory and IRS reflecting coefficients.
Fig. \ref{fig3} shows the minimum average rate of UEs obtained by different algorithms.
It is observed that the proposed algorithm achieves the best performance.
Moreover, one can see that the performance gaps between the proposed algorithm and the other algorithms are considerable and keep increasing  with the number of iterations, which verifies the effectiveness of the proposed algorithm.

\section{Conclusion}
	
In this paper, IRS-assisted and UAV-supported THz communications have been investigated.
The minimum average rate of UEs has been maximized by optimizing the UAV's trajectory optimization, the phase shift of IRS optimization, the THz sub-band allocation and the power control.
The simulation results validated the effectiveness of the proposed algorithm.

\bibliographystyle{ieeetran}
\bibliography{Reference}

\begin{thebibliography}{10}
\providecommand{\url}[1]{#1}
\csname url@samestyle\endcsname
\providecommand{\newblock}{\relax}
\providecommand{\bibinfo}[2]{#2}
\providecommand{\BIBentrySTDinterwordspacing}{\spaceskip=0pt\relax}
\providecommand{\BIBentryALTinterwordstretchfactor}{4}
\providecommand{\BIBentryALTinterwordspacing}{\spaceskip=\fontdimen2\font plus
\BIBentryALTinterwordstretchfactor\fontdimen3\font minus
  \fontdimen4\font\relax}
\providecommand{\BIBforeignlanguage}[2]{{%
\expandafter\ifx\csname l@#1\endcsname\relax
\typeout{** WARNING: IEEEtran.bst: No hyphenation pattern has been}%
\typeout{** loaded for the language `#1'. Using the pattern for}%
\typeout{** the default language instead.}%
\else
\language=\csname l@#1\endcsname
\fi
#2}}
\providecommand{\BIBdecl}{\relax}
\BIBdecl

\bibitem{9110849}
C.~{Pan}, H.~{Ren}, K.~{Wang}, M.~{Elkashlan}, A.~{Nallanathan}, J.~{Wang}, and
  L.~{Hanzo}, ``Intelligent reflecting surface aided {MIMO} broadcasting for
  simultaneous wireless information and power transfer,'' \emph{IEEE Journal on
  Selected Areas in Communications}, vol.~38, no.~8, pp. 1719--1734, 2020.

\bibitem{9090356}
C.~{Pan}, H.~{Ren}, K.~{Wang}, W.~{Xu}, M.~{Elkashlan}, A.~{Nallanathan}, and
  L.~{Hanzo}, ``Multicell {MIMO} communications relying on intelligent
  reflecting surfaces,'' \emph{IEEE Transactions on Wireless Communications},
  vol.~19, no.~8, pp. 5218--5233, 2020.

\bibitem{Chen.20198112019813}
W.~Chen, X.~Ma, Z.~Li, and N.~Kuang, ``Sum-rate maximization for intelligent
  reflecting surface based terahertz communication systems,'' in \emph{2019
  IEEE/CIC International Conference on Communications Workshops in China (ICCC
  Workshops)}.\hskip 1em plus 0.5em minus 0.4em\relax IEEE, 2019/8/11 -
  2019/8/13, pp. 153--157.

\bibitem{Pan.2020}
\BIBentryALTinterwordspacing
Y.~Pan, K.~Wang, C.~Pan, H.~Zhu, and J.~Wang, ``Sum rate maximization for
  intelligent reflecting surface assisted terahertz communications.'' [Online].
  Available: \url{https://arxiv.org/pdf/2008.12246}
\BIBentrySTDinterwordspacing

\bibitem{Chaccour.2020}
\BIBentryALTinterwordspacing
C.~Chaccour, M.~N. Soorki, W.~Saad, M.~Bennis, and P.~Popovski, ``Risk-based
  optimization of virtual reality over terahertz reconfigurable intelligent
  surfaces.'' [Online]. Available: \url{https://arxiv.org/pdf/2002.09052}
\BIBentrySTDinterwordspacing

\bibitem{Li.2020}
S.~Li, B.~Duo, X.~Yuan, Y.-C. Liang, and M.~{Di Renzo}, ``Reconfigurable
  intelligent surface assisted {UAV} communication: Joint trajectory design and
  passive beamforming,'' \emph{IEEE Wireless Communications Letters}, vol.~9,
  no.~5, pp. 716--720, 2020.

\bibitem{Han.2015}
C.~Han, A.~O. Bicen, and I.~F. Akyildiz, ``Multi-ray channel modeling and
  wideband characterization for wireless communications in the terahertz
  band,'' \emph{IEEE Transactions on Wireless Communications}, vol.~14, no.~5,
  pp. 2402--2412, 2015.

\bibitem{Han.2016}
C.~Han and I.~F. Akyildiz, ``Distance-aware bandwidth-adaptive resource
  allocation for wireless systems in the terahertz band,'' \emph{IEEE
  Transactions on Terahertz Science and Technology}, vol.~6, no.~4, pp.
  541--553, 2016.

\bibitem{Lu.2020727}
\BIBentryALTinterwordspacing
H.~Lu, Y.~Zeng, S.~Jin, and R.~Zhang, ``Aerial intelligent reflecting surface:
  Joint placement and passive beamforming design with 3d beam flattening.''
  [Online]. Available: \url{https://arxiv.org/pdf/2007.13295}
\BIBentrySTDinterwordspacing

\bibitem{Yang.2019}
\BIBentryALTinterwordspacing
Y.~Yang, S.~Zhang, and R.~Zhang, ``{IRS}-enhanced {OFDMA}: Joint resource
  allocation and passive beamforming optimization.'' [Online]. Available:
  \url{https://arxiv.org/pdf/1912.01228}
\BIBentrySTDinterwordspacing

\bibitem{Tang.2019}
\BIBentryALTinterwordspacing
W.~Tang and et~al., ``Wireless communications with reconfigurable intelligent
  surface: Path loss modeling and experimental measurement.'' [Online].
  Available: \url{https://arxiv.org/pdf/1911.05326}
\BIBentrySTDinterwordspacing

\bibitem{Wong.1999}
C.~Y. Wong, R.~S. Cheng, K.~B. Lataief, and R.~D. Murch, ``Multiuser {OFDM}
  with adaptive subcarrier, bit, and power allocation,'' \emph{IEEE Journal on
  Selected Areas in Communications}, vol.~17, no.~10, pp. 1747--1758, 1999.

\bibitem{Boulogeorgos.20186252018628}
A.-A.~A. Boulogeorgos, E.~N. Papasotiriou, and A.~Alexiou, ``A distance and
  bandwidth dependent adaptive modulation scheme for{ THz} communications,'' in
  \emph{2018 IEEE 19th International Workshop on Signal Processing Advances in
  Wireless Communications (SPAWC)}.\hskip 1em plus 0.5em minus 0.4em\relax
  IEEE, 2018/6/25 - 2018/6/28, pp. 1--5.

\end{thebibliography}
	
\end{document}